\newif\ifsubmode
\submodefalse

\newif\ifprintfig
\printfigtrue

\ifsubmode
  \documentstyle[12pt,aasms4,epsf]{article}
  \received{}
  \accepted{}
  \journalid{}{}
  \articleid{}{}
\else
  \documentstyle[11pt,aaspp4,epsf]{article}
\fi
 
\lefthead{B\"oker, van der Marel \& Vacca}
\righthead{The nuclear star cluster in IC~342}

\def\as{$^{\prime\prime}$} 
 
\def\bdm{\begin{displaymath}} 
\def\edm{\end{displaymath}} 
\def\beq{\begin{equation}} 
\def\eeq{\end{equation}} 
\def\bit{\begin{itemize}} 
\def\eit{\end{itemize}} 
\def\ben{\begin{enumerate}} 
\def\een{\end{enumerate}} 
\def\bfi{\begin{figure}[htb]} 
\def\bpfi{\begin{figure}[p]}

\def\ea{{\it et al.~}}

\newcommand{\etal}{{et al.~}}

\newcommand{\lta}{\lesssim}
\newcommand{\gta}{\gtrsim}
 
\newcommand{\kms}{\>{\rm km}\,{\rm s}^{-1}}
\newcommand{\pc}{\>{\rm pc}}
\newcommand{\kpc}{\>{\rm kpc}}
\newcommand{\Mpc}{\>{\rm Mpc}}
\newcommand{\Msun}{\>{\rm M_{\odot}}}

\newcommand{\mum}{\>{\mu {\rm m}}}

\begin{document}

\title{CO-bandhead spectroscopy of IC~342:\\ mass and age of the
nuclear star cluster\altaffilmark{1}}

\author{Torsten B\"oker,\altaffilmark{2} Roeland P.~van der Marel} 
\affil{Space Telescope Science Institute, 3700 San Martin Drive, 
       Baltimore, MD 21218}
\authoremail{boeker@stsci.edu, marel@stsci.edu}

\author{William D.~Vacca} 
\affil{Institute for Astronomy, Honolulu, HI 96822}
\authoremail{vacca@minerva.ifa.hawaii.edu}


\altaffiltext{1}{Based on observations obtained with the Infrared Telescope 
Facility, which is operated by the University of Hawaii under contract
to the National Aeronautics and Space Administration.}

\altaffiltext{2}{Affiliated with the Astrophysics Division, Space Science 
Department, European Space Agency.}

 
\ifsubmode\else
\clearpage\fi

 
\ifsubmode\else
\baselineskip=14pt
\fi


\begin{abstract} 
We have used the NASA Infra-Red Telescope Facility (IRTF) to observe
the nuclear stellar cluster in the nearby, face-on, giant Scd spiral
IC~342. From high resolution ($\lambda/\Delta \lambda = 21500$)
spectra at the $^{12}$CO (2-0) bandhead at $2.3\mum$ we derive a
line-of-sight stellar velocity dispersion $\sigma = (33 \pm 3) \kms$.

To interpret this observation we construct dynamical models based on
the Jeans equation for a spherical system.  The light distribution of
the cluster is modeled using an isophotal analysis of an HST V-band
image from the HST Data Archive, combined with new ground-based K-band
imaging. Under the assumption of an isotropic velocity distribution,
the observed kinematics imply a K-band mass-to-light ratio $M/L_K =
0.05$, and a cluster mass $M \approx 6 \times 10^6 \Msun$. We model
the mass-to-light ratio with the `starburst99' stellar population
synthesis models of Leitherer and collaborators, and infer a
best-fitting cluster age in the range $10^{6.8-7.8}$ years. Although
this result depends somewhat on a number of uncertainties in the
modeling (e.g., the assumed extinction along the line-of-sight towards
the nucleus, the IMF of the stellar population model, and the velocity
dispersion anisotropy of the cluster), none of these can be plausibly
modified to yield a significantly larger age. Also, the inferred age
is consistent with that found in our previous study based on the
near-infrared absorption line equivalent widths of the cluster
(B\"oker, F\"orster-Schreiber \& Genzel 1997).

Recent HST observations of large samples of spiral galaxies have shown
that nuclear stellar clusters are very common in intermediate to
late-type spirals. The cluster in IC 342 is more luminous than the
clusters found in most other nearby spiral galaxies. If the nuclear
stellar clusters in spiral galaxies all have a mass similar to that of
the cluster in IC 342, then stellar population synthesis models
indicate a median age for these clusters of several Gyrs. This may be
consistent with a scenario in which each spiral galaxy has only one
episode of nuclear star cluster formation. 
On the other hand, the incidence of {\it young} nuclear star
clusters may be high enough to indicate that the formation of these
clusters is a recurring phenomenon. Age and population studies for a
larger sample of galaxies are necessary to distinguish between these
scenarios, and to determine how these nuclear stellar clusters are
related to the secular evolution of their environment.

As a byproduct of our analysis, we infer that IC 342 cannot have any
central black hole more massive than $5 \times 10^5 \Msun$. This is
$\sim 6$ times less massive than the black hole inferred to exist in
our Galaxy, consistent with the accumulating evidence that galaxies
with less massive bulges harbor less massive black holes.
\end{abstract}


          
\clearpage


\section{Introduction}
\label{s:intro}

The central regions of spiral galaxies have traditionally been well
studied at $\gta 1 \kpc$ scales using ground-based
observations. Results from such studies have shown that spiral
galaxies have bulges, and these bulges diminish progressively in
prominence from early-type spirals to late-type spirals (e.g., Wyse,
Gilmore \& Franx 1997). This has formed one of the central ideas
behind the Hubble sequence classification. Recently, however,
observations with the Hubble Space Telescope (HST) have revealed that
the morphological and photometric properties of the centers of spiral
galaxies at $10$--$100 \pc$ scales are more complicated than was
previously assumed (Carollo \etal 1997; Carollo, Stiavelli \& Mack
1998; Carollo \& Stiavelli 1998). The majority of early-type spirals
(S0a--Sab) show smooth bulge-like structures with brightness profiles
that follow an $R^{1/4}$ law. Such bulges have traditionally been
viewed as `small ellipticals'. By contrast, most intermediate (Sb--Sc)
type spirals have bulges that follow an exponential profile, and are
of lower surface brightness and density than their $R^{1/4}$
counterparts. In late type spirals (Scd and later type) virtually all
bulges are of this exponential type. This transition from $R^{1/4}$ to
exponential bulges along the Hubble sequence had been inferred from
ground-based work as well (Andredakis \& Sanders 1994; Andredakis,
Peletier \& Balcells 1995; Courteau, de Jong \& Broeils 1996), but
only the HST observations revealed that a compact, photometrically
distinct nuclear star cluster is almost always present in the centers
of exponential bulges (e.g., Carollo \etal 1998; B\"oker \etal 1999).
Ground-based spectroscopic observations had already demonstrated the
existence of such clusters (e.g., Ho, Filippenko
\& Sargent 1997), but with HST they could now be resolved for the first time.
Nuclear star clusters may play an important role in the secular
evolution of intermediate and late-type spirals (Carollo 1999). In
order to better understand the origin of these clusters and the
influence they have on their surroundings, more detailed information
is now needed on their properties, including their masses and the ages
of their stellar populations.

In this paper we present a detailed study of the nuclear star cluster
in IC 342, a nearby, face-on giant Scd spiral.  The distance to IC
342, $D=1.8 \Mpc$ (1\as = 8.7 pc) as advocated by McCall (1989), makes
it the nearest giant spiral after M31 and M33 (although it has
recently been suggested that the distance may be larger, $\sim 3$--$4
\Mpc$; see \S\ref{ss:considerations} below). As a result of its small 
distance and face-on orientation, the nuclear star cluster in IC 342
is better resolved, apparently brighter, and less obscured by disk
material than in nearly all other spiral galaxies. The only difficulty
with observations of IC 342 is its proximity to the Galactic plane,
which results in several magnitudes of foreground extinction (McCall
1989; Madore \& Freedman 1992). This explains the paucity of previous
optical studies of this nearby galaxy (as well as its absence from the
Messier and NGC Catalogs). However, it is not a major drawback for
infrared studies; the extinction in the K-band is nearly 10 times
smaller than in the V-band (e.g., Rieke \& Lebofsky 1985), and
near-infrared observations can therefore almost completely penetrate
the dust.

Because of its enhanced star formation rate (e.g., Becklin \etal
1980), IC 342 has previously been a popular target for infrared and
sub-mm observations.  In B\"oker, F\"orster-Schreiber \& Genzel (1997,
hereafter BFG97) we obtained medium-spectral resolution
two-dimensional near-infrared spectroscopy in the H and K-bands for
the central 12\as\ of IC~342. While part of this study focused on the
$\sim$ 7\as\ diameter ring of molecular gas and
star-formation in IC 342 (Turner \& Ho 1983; Ishizuki \etal 1990), we
also addressed the nature of the star cluster seen at the very center
of the galaxy. The observations used to constrain the latter were: (i)
the equivalent widths of several prominent absorption lines; and (ii)
an estimate of the mass-to-light ratio ($M/L$) of the cluster. From
stellar population synthesis models we inferred that the cluster is
presumably young ($\sim 10^7$ years) and dominated by a population of
red supergiants.

Our previous analysis of the nuclear cluster remained somewhat
tentative, however. In particular, the $M/L$ estimate of the cluster was based
on the molecular gas rotation curve obtained by Turner \& Hurt
(1992). There are several reasons why the resulting $M/L$ could be highly
uncertain.  For example, hydrodynamical forces may cause the gas
rotation speed to deviate from the circular velocity, and the limited
spatial resolution ($\sim$ 5\as ) of the gas kinematic observations
may yield an underestimate of the rotation velocity.  We therefore set
out to obtain a more direct and more accurate determination of the
$M/L$.

The velocity dispersion of the nuclear star cluster in IC~342 provides
a direct measure of its mass, and hence its $M/L$. In this paper we
report on high-resolution spectroscopy of the cluster around the
infrared $^{12}$CO~(2-0) bandhead at $2.2935 \mum$. This spectral
region is relatively insensitive to dust absorption, and has been
demonstrated to be well-suited for the determination of the velocity
dispersion of stellar systems (e.g., Gaffney, Lester \& Doppmann
1995). To facilitate the interpretation of the spectroscopic data we
also obtained a new K-band image of IC~342.

The outline of the paper is as follows. In \S\ref{s:data} we describe
the observations, the data reduction and the kinematic analysis from
which we infer the velocity dispersion of the nuclear star cluster. In
\S\ref{s:models} we analyze a V-band image of IC~342 from the HST Data
Archive, as well as the new K-band ground-based image.  We then
present dynamical models for the cluster from which we derive the
mass-to-light ratio, and we interpret the results with
stellar-population synthesis models. In \S\ref{s:summary} we we
present some concluding remarks, and we discuss the results of our
analysis in the context of our general understanding of the properties
and evolution of spiral galaxy nuclei and bulges.

\section{Observations, data reduction and kinematic analysis}
\label{s:data}

\subsection{K-band imaging}
\label{ss:Kimage}

We obtained a K-band image of the central region of IC~342 at the NASA
Infrared Telescope Facility (IRTF) on Mauna Kea, Hawaii, on the night
of 29 November 1998 (UT) using the NSFCam 256x256 InSb array camera
(Shure \etal 1994) with the 0.3\as /pixel plate scale.  Conditions
were photometric and the seeing was estimated to be $\sim$
0.5\as --0.6\as\ FWHM. Seven images of IC 342, each consisting of 50
coadds of individual frames with integration times of 0.4s, were
acquired at different positions on the array. Exposures of an offset
sky position 200\as\ east of the galaxy were obtained immediately
after each galaxy image and with identical integration times. The
offset frames were combined to construct a sky image and a sky flat,
which were used to process (sky subtract and flat-field) the galaxy
images. The dark current is negligible for these short integration
times and did not have to be subtracted from the sky frames before
constructing the flat field. The individual galaxy images were then
shifted to a common reference frame (registered) and combined into a
single mosaic. The final image is shown in Figure~\ref{f:images}a. The
stellar images on the mosaic reveal a slight degradation of the image
quality by the mosaic process to about 0.7\as\ FWHM.

We also acquired images of several UKIRT faint standard stars during
the night in order to derive photometric calibrations. The images were
obtained and reduced in a manner similar to that for the galaxy
images. Separate sky frames were not acquired, however; rather, the
standard star fields were dithered on the array and were used to
generate the sky frames and flat fields.  Photometry was performed
using DAOPHOT in IRAF with a 12 pixel radius aperture and the
photometric zero point and K-band extinction coefficient were
determined. These coefficients were applied to the aperture photometry
of the K-band mosaic discussed in \S\ref{ss:surfbr} below.

\subsection{Spectroscopic Observations and Data Reduction}
\label{ss:observations}

We obtained high resolution spectra of IC 342 centered on the CO
bandhead with the CSHELL spectrograph on the IRTF, during parts of the 
nights of 1998 October 14, November 29, and December 31. CSHELL is a
long-slit spectrograph which uses a 31.6 lines/mm
echelle with narrow-band circular variable filters (CVFs) that isolate a
single order (Tokunaga \etal 1990; Greene \etal 1993). It operates
from $1.08$--$5.6 \mum$. The detector is a $256
\times 256$ pixel Hughes SBRC InSb array with a pixel scale of 0.2\as\ 
in the spatial direction and $2.7 \kms$ in the wavelength direction;
the corresponding free spectral range is $691 \kms$. The slit length
is 30\as\, or 150 spatial pixels. The grating was set to position the
CO bandhead on the array center, and the corresponding order was
isolated with the circular variable filter. The systemic velocity of
IC 342 is only $35 \kms$ (Turner \& Hurt 1992) and we did not
compensate for it in the grating positioning. The seeing during the
observations varied from 0.7--1.2\as\ (FWHM). All observations were
obtained with a 1.0\as\ slit width, which corresponds to 5 pixels on
the detector, or $13.5\kms$ in the spectral direction. The resulting
resolving power is $R \equiv \lambda / \Delta \lambda = 21,500$, with
a Gaussian dispersion of the instrumental line-spread-function of
$\sigma_{\rm instr} = 5.5 \kms$ (as measured from arc-lamp spectra).

Our observing procedure was the same on each of the three
nights. After moving the telescope to IC 342, we obtained an image of
the slit using an internal calibration lamp in order to determine the
position of the slit on the detector. We then moved the slit out of
the beam and acquired a direct image of IC 342 through the CVF
(centered at $2.3\mum$) using the imaging mode of CSHELL (in which a
plane mirror substitutes for the grating).  This image was used to
accurately position the $2.3\mum$ peak of IC 342 on the slit. Another
image of the slit was acquired to insure that the object was
positioned correctly. The adopted strategy of positioning the target
using its near-IR light avoids possible pointing errors due to
differential atmospheric refraction.  Subsequent guiding was usually
done using the visible light from the target, as observed with a CCD
detector that is mounted behind a dichroic in the beam.

The CSHELL slit was rotated to a position angle of $0$ degrees (i.e,
North-South) and the galaxy spectra were acquired at two different
positions along the slit, separated by $\sim$ 12\as . Exposures of the
nucleus of IC 342 were alternated with equal length exposures of an
offset sky position (blank field) located 100--200$^{\prime\prime}$
east. In this manner, we obtained 17 galaxy spectra during the course
of the three nights with typical exposure times of 240--300s each, for
a total of 4740s on-source. Flat field frames (obtained using the
internal continuum lamp) and dark frames were acquired immediately
after the exposures of IC 342 and at the same sky position.  We also
observed several A0 stars to allow correction for telluric absorption
features, and four giant stars of spectral types K and M, for
comparison to the galaxy spectrum in the kinematic analysis.  At
several times during each night we obtained spectra of internal Ar and
Kr arc lamps, for use in wavelength calibration.

The data processing was carried out with standard two-dimensional spectral
reduction routines in IRAF. We built flat-fields from the normalized,
dark-subtracted continuum lamp spectra. These were used to flat-field
all external-target exposures. The dark and sky contributions were
then removed from each galaxy exposure by subtraction of the
associated blank-field exposure. Before doing so, we had to scale each
blank-field exposure to the sky level observed in the associated
galaxy spectrum, to correct for temporal sky brightness variations.
For the star observations we did not obtain blank field spectra, but
instead, each star was observed at two different positions along the
slit. Dark- and sky-contribution removal for these observations was
achieved by differencing the two spectra for each star, with
additional scaling where necessary. All star and galaxy exposures were
then rebinned simultaneously in two dimensions, to calibrate the
scales in both the wavelength and spatial directions; the wavelength
rebinning was done linearly in $\log \lambda$, as appropriate for the
kinematic analysis. The wavelength scale was calibrated using the
known wavelengths of the emission lines in the arc lamp exposures,
while the spatial tilt of the spectra on the detector was removed
using the observations of stars at the various positions along the
slit.
 
To correct the resulting calibrated spectra for telluric absorption
features, we used observations of A0 stars at similar airmasses as
those of the galaxy and velocity standards. The spectra of these stars
are virtually featureless over the small spectral range available with
CSHELL.  Division of their spectra by the normalized spectrum of a
black-body with a temperature of 9500$\>$K, gives a good measure of
the atmospheric absorption. The results were divided into the galaxy
and template star spectra. All individual galaxy and star exposures
were then co-added with cosmic-ray removal, and all resulting spectra
were subsequently collapsed in the spatial direction to yield final
one-dimensional spectra. In the latter step we co-added the central
5--7 spatial rows (1.0--1.4\as ) of each two-dimensional spectrum
(corresponding roughly to the seeing FWHM for the observations). The
nuclear cluster of IC 342 is smaller than the seeing disk of the
spectroscopic observations (cf.~\S\ref{ss:surfbr} and
Figure~\ref{f:images}b below), and collapsing its spectrum spatially
therefore does not lead to loss of information. We did not perform a
flux calibration on any of the spectra, since this is not required for
the kinematic analysis.

Figure~\ref{f:spectra} shows the final reduced spectrum of the nuclear
cluster of IC 342 (top), as well as the spectra of the four template
stars that we observed for the kinematic analysis (bottom). The
galaxy spectrum has a signal-to-noise ratio $S/N \approx 19$ per
spectral pixel, and that of the template spectra is several times
higher. Whereas the $S/N$ of the galaxy spectrum is limited primarily
by sky-noise, the $S/N$ of the template spectra is limited primarily
by the noise in the observations of the atmospheric
calibrators. Residual telluric absorption due to imperfect matches in
airmass are present at levels up to 10 per cent. However, this is not
critical for the kinematic analysis, since broadening of the
template spectra with the modeled velocity distribution of the galaxy
automatically smoothes away possible residual telluric features. We
verified explicitly that the inferred velocity dispersion of IC 342
(see below) does not change significantly if galaxy spectra from
individual nights are analyzed, or if the galaxy spectra are modeled
with different template spectra.  The results of this exercise
confirmed that residual telluric absorption is not an important
limiting factor in the analysis.

\subsection{Kinematic analysis}
\label{ss:kinanalysis}

Methods for stellar kinematic analysis assume that an observed
galaxy spectrum is the convolution of a suitable template spectrum
with the galaxy's line-of-sight velocity distribution. Various methods
exist to determine the best-fitting velocity distribution for given
galaxy and template spectra (reviewed in, e.g., Binney \& Merrifield
1998). Here we have used the `Gauss-Hermite Pixel Fitting Software' of
van der Marel (1994)\footnote{This software is available at
http:/$\!$/www.stsci.edu/$\sim$marel/software.html.}. It parameterizes
the velocity distribution as a Gauss-Hermite series (van der Marel \&
Franx 1993), and determines the best-fitting parameters using direct
chi-squared minimization in pixel-space. The formal errors in the fit
parameters follow from the shape of the chi-squared surface near its
minimum. This method has been well tested, and its results agree with
those from other methods for extracting stellar kinematics from galaxy
spectra.

Our spectroscopic data on IC 342 have
insufficient $S/N$ to determine the higher-order
Gauss-Hermite moments of the velocity distribution, and we have
therefore restricted the analysis to the determination of the
parameters $(\gamma,V,\sigma)$ of the best-fitting Gaussian broadening
function. The quantity $V$ measures the velocity difference between
the galaxy and the template star. This can be used to infer the
systemic velocity of the galaxy (if the template star velocity is
known a priori), but this is of little interest here. The quantity
$\gamma$ is the `line-strength' of the galaxy, or equivalently, the
average ratio of the absorption line equivalent widths in the galaxy
and those in the template star. For the four template stars that we
have observed we find $\gamma = 1.21 \pm 0.05$. This result, in
principle, yields some information on the stellar population of the
nuclear star cluster in IC 342. However, stellar population analyses
from absorption line strengths are more suitably performed using lower
spectral-resolution observations of a wider range of absorption
systems, as was done in BFG97. We therefore restrict
ourselves in the present paper to the quantity $\sigma$, which
measures the line-of-sight velocity dispersion of the stars in the
cluster. The chi-squared fit to the spectrum yields $\sigma = (33 \pm 3)
\kms$. The heavy solid curve in Figure~\ref{f:spectra} shows the 
corresponding best fit to the galaxy spectrum. For comparison, a heavy
dashed curve shows the (unacceptable) fit when the velocity dispersion
of the nuclear cluster is fixed to $\sigma = 62 \kms$, whereas the
stellar template spectra themselves correspond to $\sigma = 5.5 \kms$
(the instrumental dispersion). It is clear that the observed gradient
at the (2--0) $^{12}$CO bandedge provides a sensitive measure of the
stellar velocity dispersion broadening (as discussed previously in
detail by, e.g., Gaffney, Lester \& Doppmann 1995). Differences in the
inferred $\sigma$ obtained with different template stars were found to
be at the level of $\pm 2 \kms$. This is less than the random error on
the inferred dispersion, and template mismatch is therefore not an
important source of error in the analysis. In particular, we found no
significant difference in the inferred $\sigma$ for either a red
supergiant or a red giant as the template star.

\section{Modeling and Interpretation}
\label{s:models}

\subsection{HST surface brightness profile and ground-based photometry}
\label{ss:surfbr}

To properly interpret the velocity dispersion of the nuclear cluster,
we also need to measure its spatial structure. We searched the HST
Data Archive for images of IC 342, and found that broad-band images
have been obtained with the Second Wide Field and Planetary Camera
(WFPC2) in the filters F555W, F606W, F675W and F814W. Unfortunately,
in all images the central parts of the nuclear cluster were found to
be saturated.  Of the images obtained with the PC chip which, with a
pixel scale of 0.046\as , yields the highest spatial resolution, the
ones least affected by saturation are two 80s exposures in the F606W
filter. We concentrate on these images here. They were obtained by
Illingworth and collaborators as part of Cycle~4 program GO-5446.  We
downloaded the pipeline-calibrated images, registered them, and
combined them with cosmic-ray rejection. The resulting image of the
nuclear region is shown in Figure~\ref{f:images}b.

Figure~\ref{f:surfbr} shows the major axis surface brightness profile
of the nuclear star cluster as inferred by ellipse fitting to the
observed isophotes. 
The fitting was performed using software found in the IRAF
package STSDAS, which is based on the algorithm of Jedrzejewski
(1987). Image counts were transformed to V-band magnitudes using the
procedures described by Holtzman \etal (1995). Pixels at $r \leq$
0.08\as\ are saturated at a level that corresponds to a surface
brightness of $12.48$ mag/arcsec$^2$, which causes an apparent flattening
of the surface brightness profile at these radii. The figure also
shows the PSF profile. Outside the radii affected by saturation, the
galaxy profile is clearly more extended than the PSF. The nuclear
cluster is therefore resolved.

For use in the dynamical modeling we fitted the observed surface
brightness profile with a so-called `nuker-law' parametrization (Lauer
\etal 1995; Byun \etal 1996). Two fits are shown as heavy solid curves in
Figure~\ref{f:surfbr}. The lower curve is a fit to the data at $r \leq$
0.3\as , including the saturated pixels. The upper curve is a fit over
the same range, but excluding the saturated pixels at $r \leq$ 0.08\as .
The former model has a homogeneous core, whereas the latter model has
a power-law surface brightness cusp similar to those typically found
in the centers of elliptical galaxies (e.g., Faber \etal 1997). Both
provide an acceptable fit to the available data.

Inspection of the ground-based K-band image (Figure~\ref{f:images}a)
at different stretches reveals that it shows most of the same features
as the HST V-band image (Figure~\ref{f:images}b), including the two
clusters $\sim$ 2\as\ to the North and South-South-West of the nuclear
cluster. However, the lower spatial resolution of the K-band image
makes all features more difficult to discern. The morphology of the
diffuse background light in the K-band image is more symmetric than in
the HST image, due to the smaller influence of dust absorption in the
K-band.

From the NSFCam image we measured a total magnitude $K=10.59$ in a
1.2\as\ diameter aperture centered on the nuclear cluster. To obtain
the optical-infrared color of the nuclear cluster we convolved the HST
image with a 0.7\as\ FWHM Gaussian, and found that $V=15.19$ in a
1.2\as\ diameter aperture at the same position. This implies
$V-K=4.60$ for the average color of the nuclear cluster. The
saturation of the HST image in the central few pixels does not provide
a significant uncertainty in this measurement; use of the cusped
nuker-law profile shown in Figure~\ref{f:surfbr} (instead of the
actual image) yields a value that is only modestly different, $V-K =
4.54$. In the remainder we adopt the average of these measurements,
$V-K = 4.57$. In BFG97 we presented a K-band image of IC 342 obtained
in 1\as\ FWHM seeing. The K-band photometry and $V-K$ color inferred
from our new image are completely consistent with the absolute
photometry from that paper.

\subsection{Dynamical modeling}
\label{ss:dynmodel}

We use our data to estimate the mass-to-light ratio $M/L_K$ of the
nuclear stellar cluster in the K-band. The K-band light of the cluster
is less sensitive to dust absorption than the V-band light, and
$M/L_K$ can therefore be more accurately determined than $M/L_V$.
This is important, because the sightline towards IC 342 suffers from a
considerable amount of extinction. In BFG97 we estimated $A_V = 4$,
based on the observed $H-K$ color of the nucleus. With the Galactic
interstellar extinction law of Rieke \& Lebofsky (1985) this implies
$A_K = 0.45$, which we will assume for the moment. In
\S\ref{ss:popsynthesis} we will discuss the extinction and its
influence on our results in more detail.

We assume that the observed $V-K$ color of the cluster does not vary
significantly over its (small) spatial extent.  The combination of the
V-band surface brightness profile inferred from the HST data, combined
with $V-K=4.57$ and $A_K = 0.45$ then yields the intrinsic
unextinguished K-band surface brightness profile.  The analysis then
proceeds as in van der Marel (1994). The surface brightness profile is
deprojected under the assumption of spherical symmetry, and the
three-dimensional velocity dispersion profile $\sigma(r)$ is
calculated by solving the Jeans equation for a spherical isotropic
system. The results are then projected along the line of sight and
convolved with the IRTF observational setup, to yield a prediction for
the observed velocity dispersion. This dispersion scales as $\sigma
\propto \sqrt{M/L}$, and we determine the $M/L$ that produces the
observed $\sigma = 33 \kms$. For the homogeneous-core model in
Figure~\ref{f:surfbr} we infer $M/L_K = 0.050$, and for the cusp model
$M/L_K = 0.044$ (all mass-to-light ratios are given in units of
$\Msun/L_{\odot,K}$). Both results assume a distance to IC 342 of $D =
1.8 \Mpc$ (McCall 1982; Madore \& Freedman 1992; more on this in
\S\ref{ss:considerations} below). The models show that the inferred
$M/L$ is not particularly sensitive to the assumed behavior of the
surface brightness profile at small radii, so the saturation of the
HST images is not a great limitation in our analysis. In the following
we adopt the average of the two inferred values as our best estimate:
$M/L_K = 0.047 \pm 0.010$, where the error is the formal error in the
$M/L_K$ determination due to the formal error in the velocity
dispersion $\sigma$. 

The mass-to-light {\it ratio} of the cluster is well-determined by the
observations. However, this is less true for either the {\it total}
light or the {\it total} mass of the cluster, given that the cluster
blends smoothly into a diffuse background (cf.~Figure~\ref{f:surfbr}).
We have used two approaches to obtain rough estimates for these
quantities. In the first approach, we start from the observed
$K=10.59$ in a 1.2\as\ diameter aperture. If the cluster were a point
source then in 0.7\as\ FWHM Gaussian seeing, the flux contained in a 1.2\as\
aperture would be 86\%. This is a strict upper limit, because the cluster is
more extended than a point source. However, the HST data do show the
cluster to be very compact, so 86\% may be a reasonable
estimate. Taking into account an extinction of $A_K = 0.45$, we find
$K = 9.98$ for the intrinsic cluster luminosity. At $D=1.8\Mpc$
this corresponds to $7.6 \times 10^7 L_{\odot,K}$. Multiplication by
$M/L_K = 0.047$ yields $M = 3.6 \times 10^6 \Msun$. In the second
approach, we have calculated the total luminosity (integrated out to
infinity, and corrected for extinction) of the spherical model
clusters that generate the model surface brightness profiles shown as
solid curves in Figure~\ref{f:surfbr}. This yields $1.8 \times 10^8
L_{\odot,K}$ for both the core and the cusp model. Multiplication by
$M/L_K = 0.047$ yields $M = 8.4 \times 10^6 \Msun$. It is likely that
these estimates bracket the true cluster mass, which is therefore $M
\approx (6.0 \pm 2.4) \times 10^6 \Msun$.

\subsection{Population synthesis}
\label{ss:popsynthesis}

The inferred $M/L_K$ of the nuclear stellar cluster yields a
constraint on its age. Figure~\ref{f:masstolight} shows the relation
between $M/L_K$ and age for several model stellar clusters, as
calculated with the `Starburst99' software of Leitherer
\etal (1999)\footnote{This software is available at
http:/$\!$/www.stsci.edu/science/starburst99/.}. Curves are shown for
the same three standard assumptions for the initial mass function
(IMF) used by Leitherer et al. All curves assume solar metallicity,
and star formation in a single, instantaneous burst (continuous star
formation models are discussed and dismissed below). Curves for
different metallicities do not look markedly different (Leitherer
\etal 1999). A general property of all these models is that $M/L$
increases strongly with time after $\sim 10^{7.5}$ years. The low
value of $M/L_K$ inferred for the cluster in IC 342 therefore
indicates that it must be rather young, $\log({\rm age}) \lta 7.7$.

The three IMF models in Figure~\ref{f:masstolight} all have a
lower-mass cutoff at $M_{\rm low} = 1 \Msun$. Stars with masses lower
than this contribute little to the light output of clusters younger
than $\sim 1$ Gyr. However, low-mass ($M \lta 1\Msun$) stars may
constitute a significant fraction of the total mass of the cluster.
The curves in Figure~\ref{f:masstolight} should therefore be regarded
as lower limits. Given our observed constraint on the $M/L_K$ of the
cluster, any addition of low-mass stars to the IMF will lower the
inferred age. Since it is our aim here to confirm that the cluster is
young, the neglect of low-mass stars is conservative.

The observed optical-infrared color of the nuclear cluster, $V-K =
4.57$, provides an additional constraint on stellar population models
for the nuclear cluster. However, this constraint is more difficult to
interpret, because of the large extinction-induced reddening in
IC~342. The extinction towards IC 342 (galactic + internal) has
previously been estimated to be $A_V \approx 2.5$ magnitudes (McCall
1982; Madore \& Freedman 1992). These estimates were obtained for
regions between $0.5$ and 10 arcmin from the galaxy center. However, the
dust absorption towards IC 342 is very patchy, and variable on scales
as small as a fraction of an arcsecond (cf.~Figure~\ref{f:images}b).
These previous extinction measurements can therefore not be
confidently used to estimate the extinction of the light from the
nuclear star cluster. Instead, we can estimate the extinction from the
observed color $(V-K)_{\rm obs}$, if we assume that the intrinsic
color $(V-K)_{\rm int}$ is known from stellar population models.  One
has $(V-K)_{\rm obs} = (V-K)_{\rm int} + (A_V - A_K)$. For the
Galactic interstellar extinction law of Rieke \& Lebofsky (1985), $A_K
= 0.112 A_V$. Hence, $A_K = 0.112 A_V = 0.126 [ (V-K)_{\rm obs} -
(V-K)_{\rm int} ]$. Figure~\ref{f:vminusk} shows $(V-K)_{\rm int}$ for
the same stellar population models as in Figure~\ref{f:masstolight}.
The predicted values range from $-0.42$ to $3.35$. This constrains
$A_K$ to the range $0.15$--$0.63$, independent of the cluster age.

A more complete treatment involves simultaneous fitting of the
observed $M/L_K$ and $V-K$, as function of the two independent
variables age and $A_K$. This is done in Figure~\ref{f:AKage}. Each
panel corresponds to one of the three population models in
Figures~\ref{f:masstolight} and~\ref{f:vminusk}. The dotted curve in
each panel shows, as function of cluster age, the value of $A_K$ for
which the predicted $V-K$ of the population model matches the
observations. The solid curve in each panel shows, again as function
of cluster age, the value of $A_K$ for which the predicted $M/L_K$ of
the population model matches the observations (i.e., $M/L_K = 0.047
\times 10^{-0.4 (A_K - 0.45)}$, cf.~\S\ref{ss:dynmodel}). The 
points in the panels where the solid and dotted curves intersect
indicate combinations of age and $A_K$ that fit both the observed
color and mass-to-light ratio. The allowed combinations have ages in
the range $10^{6.8-7.8}$ years, and $A_K$ in the range
$0.37$--$0.57$. The latter range includes the value $A_K \approx 0.45$
(i.e., $A_V = 4.0$) from BFG97 that was adopted in
\S\ref{ss:dynmodel}.

In BFG97 we presented the equivalent widths of three prominent
absorption lines in the H- and K-band spectra of the nucleus of IC
342, namely Si($1.59\mum$), CO($\nu =$6--3) at $1.62\mum$ and CO($\nu
=$2--0) at $2.29\mum$. From these measurements we argued previously
that the nucleus must be quite young. For completeness, we compare in
Figure~\ref{f:eqwidths} the observed equivalent widths to the
predictions for the same stellar population models as in
Figures~\ref{f:masstolight} and~\ref{f:vminusk}. The cluster ages
implied by each of these measurements are fully consistent with those
inferred from the observed $M/L_K$ and $V-K$.

Figures~\ref{f:masstolight}--\ref{f:eqwidths} all show the evolution
of stellar populations that formed in an instantaneous burst at $t=0$.
However, the star formation rate $S(t)$ of the cluster in IC 342 need
not necessarily have been a delta-function
$\delta(t)$. Figure~\ref{f:contburst}a shows the predicted $M/L_K$ for
a model that has had a constant star-formation rate since $t=0$ (also
from Leitherer \etal 1999). In such a model $M/L_K$ increases
monotonically with cluster age after $\sim 10^7$ years, but much more
gently than for an instantaneous burst model. As a consequence, the
observed $M/L_K$ places a less stringent constraint on the cluster
age. However, models with constant star-formation rate are ruled out
for IC 342, because of the large observed equivalent width of the CO
bandhead at $2.29\mum$. The measured value is so high that it can only
barely be fit by any instantaneous burst population
(cf.~Figure~\ref{f:eqwidths}c). Any extended star formation history
mixes populations of different ages, and therefore predicts a lower
maximum CO bandhead strength. This is illustrated in
Figure~\ref{f:contburst}b, which shows the predicted CO bandhead
strength for a continuous burst model. The models cannot reproduce the
observed value for any age. This argument can also be turned around:
since an instantaneous burst model fits the observed value only for a
range of ages that is $\sim 5 \times 10^6$ years, the star formation
in IC 342 must have occurred over a timescale shorter than this.
Since the suspected cluster age ($10^{6.8-7.8}$ years) exceeds this
timescale by a large factor, the use of instantaneous burst models for
IC 342 is justified.

\subsection{Uncertainties and additional considerations}
\label{ss:considerations}

In this section we elaborate on the analysis presented in the
preceding sections, and we highlight some of the sources of
uncertainty in our conclusions.

Our dynamically inferred mass-to-light ratio for IC 342 is inversely
proportional to the assumed galaxy distance. Historical distance
estimates for IC 342 have ranged 1.5 to $8\Mpc$
(Ables 1971; Sandage \& Tammann 1974), due to differences in the
assumed extinction. Detailed work by McCall (1989) convincingly supported
a short distance of $1.8 \Mpc$, and his results were supported by
subsequent work (e.g., Madore \& Freedman 1992; Karachentsev \&
Tiknonov 1993). However, IC 342 is a member of a nearby group that
also includes Maffei~1 and~2, and at least 9 other galaxies (e.g.,
Tully 1988; Karachentsev \etal 1993). The surface brightness
fluctuation method was used to determine a distance to Maffei 1 of
$4.2 \pm 0.5 \Mpc$ (Luppino \& Tonry 1993), while the Tully-Fisher
relation was used to determine distances to the group members NGC 1560
and UGCA 105 of $3.5 \pm 0.7$ and $3.8 \pm 0.9 \Mpc$, respectively
(Krismer, Tully \& Gioia 1995). The latter authors used this to argue
for a distance of $\sim 3.6 \Mpc$ for the entire IC 342/Maffei
group. If this is true, than our $M/L_K$ for IC 342 must be reduced by
a factor~2. Figure~\ref{f:masstolight} shows that this would imply an
even younger age than inferred above.

It is likely that some uncertainty remains in our results due to dust
extinction towards IC 342. However, we have minimized these
uncertainties by performing our spectroscopic observations in the
near-infrared. Also, the luminosity of the nucleus, which enters into
the denominator of the inferred mass-to-light ratio, is based on our
K-band image. The V-band HST image of Figure~\ref{f:images}b was used
in the mass-to-light ratio determination only to obtain the spatial
distribution of the light, but not its total flux (which in the V-band
is very sensitive to extinction). Nonetheless, even in the K-band
there is some extinction, which affects the inferred $M/L_K$ and hence
the implied age.  
The use of foreground screen models for the extinction law seems to be
justified by the face-on orientation of IC~342 and the fact that most
of the extinction is due the the disk of the Milky Way. Nonetheless,
part of the extinction is likely to be intrinsic to IC 342, and may
follow another extinction law. The extinction law for starburst
galaxies inferred by Calzetti (1997) does not have the same $A_K/A_V$
as the Galactic value from Rieke \& Lebofski (1985) used in
Figure~\ref{f:AKage}. However, errors in the inferred $A_K$ for IC~342
can not cause the inferred $M/L_K = 0.047$ to be too low by more
than a factor $10^{0.4 \times 0.45} = 1.51$ (corresponding to the case
$A_K = 0$). Figure~\ref{f:masstolight} shows that this is too small to
significantly affect the estimated age.

The fact that the HST image on which we base our analysis is saturated
at small radii causes some uncertainty. However, as demonstrated in
\S\ref{ss:dynmodel}, the inferred $M/L$ differs by only 12\% for the
homogeneous-core and the cusp model in Figure~\ref{f:surfbr}. In
principle, the real cluster profile could be even steeper at small
radii than in our cusp model; it could be as steep as the PSF.  This
would decrease the $M/L$ from the derived value, because an
increased cusp slope implies that there is more light (higher $L$),
while the cluster also becomes more compact, which decreases $M$ ($M
\propto \sigma^2 r$ according to the virial theorem). Therefore, a steeper 
cusp slope of the cluster would imply an even younger age than
derived above.
 
So far we have assumed that the velocity distribution of the nuclear
stellar cluster is isotropic. This is plausible, given that two-body
relaxation in a cluster as dense as this will tend to remove any
velocity anisotropy. However, if the cluster is indeed very young,
then two-body relaxation may not have had sufficient time to
operate. Therefore, we have also constructed models with anisotropic
velocity distributions, to assess the influence of this on the
inferred $M/L$. As in \S\ref{ss:dynmodel}, we have constructed the
models by solving the Jeans equations. We find that the inferred $M/L$
increases monotonically with the ratio $\sigma_t / \sigma_r$ of the
velocity dispersions in the tangential and radial directions (assumed
to be independent of radius). However, the dependence on anisotropy is
not large. A rather extreme radial anisotropy of $\sigma_t / \sigma_r
= 1/3$ decreases $M/L$ by 10\% with respect to the isotropic model,
whereas a similarly extreme tangential anisotropy $\sigma_t / \sigma_r
= 3$ increases $M/L$ by 10\%. From Figure~\ref{f:masstolight} it is
clear that these uncertainties in $M/L_K$ due to the unknown velocity
dispersion anisotropy are small enough to not significantly affect the
inferred age. We note also that the nuclear star cluster may not be
spherical. However, deviations from spherical symmetry are expected to
influence the inferred $M/L$ only at the same (insignificant) level as
does velocity anisotropy.  We have therefore not attempted to
construct axisymmetric dynamical models, which are significantly more
complicated than the spherical models used here (e.g., Cretton \etal
1999).

The observed velocity dispersion of IC 342 also yields an upper limit
on the mass of a possible central black hole. The largest possible
black hole mass is obtained under the assumption that the stars
themselves contribute no mass at all. In this case our isotropic Jeans
models indicate that the observed $\sigma=33 \kms$ requires $M_{\rm
bh} = 5.0 \times 10^5 \Msun$ for the homogeneous-core model in
Figure~\ref{f:surfbr} and $M_{\rm bh} = 3.3 \times 10^5 \Msun$ for the
cusp model.  Therefore, any black hole in IC 342 must be less massive
than these limits.

It is interesting to note in this context that the total luminosity of
IC 342 is not much different than that of our own Galaxy (McCall
1979). However, our Galaxy harbors a black hole of $3
\times 10^6 \Msun$ (Eckart \& Genzel 1997; Ghez 1998). IC 342 is an
Scd spiral whereas our Galaxy is an Sbc spiral, so this is consistent
with the accumulating evidence that galaxies with less massive bulges
harbor less massive black holes (e.g., Kormendy \& Richstone 1995; van
der Marel 1999). If IC 342 does have a central black hole, then the
$M/L$ of the stellar population would have to be smaller than inferred
above, implying a younger age (cf.~Figure~\ref{f:masstolight}).

\section{Discussion and Conclusions}
\label{s:summary}

We have obtained high-resolution CO-bandhead spectra of the nuclear
stellar cluster in IC~342 with the IRTF, from which we derive a
velocity dispersion $\sigma = (33 \pm 3) \kms$. To interpret this
result we have constructed dynamical models for the cluster in which
the light distribution is derived from an HST V-band image from the HST
Data Archive, combined with new ground-based K-band imaging. The
models yield $M/L_K \approx 0.047$ for the mass-to-light ratio of the
cluster, and $M \approx 6 \times 10^6 \Msun$ for the cluster
mass. Stellar population models for the observed $M/L_K$, the observed
$V-K$ color, and the observed near-infrared absorption line equivalent
widths from BFG97 all consistently indicate that the cluster age is in
the range $10^{6.8-7.8}$ years. The K-band extinction towards the
nucleus is inferred to be in the range $0.37$--$0.57$. A variety of
minor uncertainties in our observations and analysis cannot
significantly alter these conclusions.

In \S\ref{s:intro} we discussed the recent changes in our
understanding of the structural properties of the centers of spiral
galaxies. We now know that many/most intermediate and late type
spirals have exponential bulges with nuclear stellar clusters in their
centers. Carollo (1999) suggests that both these components could have
formed through secular evolution within pre-existing disks. The
nuclear stellar clusters in particular could have formed as a result
of gas flow to the nucleus, initiated by torques from a nuclear
bar. Once formed, the nuclear stellar clusters observed in spirals are
massive enough to disrupt any existing nuclear bar, and also to
prevent any new bars from forming.  In this picture, the formation of
a nuclear stellar cluster itself shuts off any further gas flow into
the center. As a result, nuclear cluster formation happens only
once. One consequence of this scenario is that the nuclear clusters in
spiral galaxies would have to display a large range of ages, given
that at least some spiral galaxies are several Gyrs old (as structural
entities), and assuming that we are not viewing all spiral galaxies at
a special epoch. Carollo (1999) finds that the most luminous nuclear
star clusters ($M_V \lta -12$) are found in galaxies that display a
HII-region-like or AGN-like spectrum at ground-based resolution,
whereas the less luminous clusters ($M_V \gta -12$) do not generally
show these characteristics in their spectra. Although the mass
function of nuclear clusters is as of yet unconstrained, and
although dust extinction and non-thermal emission are likely to affect 
their optical light, it is plausible that, in general, bright clusters 
are younger than faint clusters. This is
consistent with the fact that typical stellar populations fade by several
magnitudes in V as they age (Bruzual \& Charlot 1993). 
IC~342 provides a unique opportunity to break the age-mass degeneracy,
since it is the one galaxy for which an explicit age determination is now
available.

Figure~\ref{f:carolsamp} presents a histogram of nuclear stellar
cluster luminosities for the combined galaxy sample of Carollo,
Stiavelli \& Mack (1998), kindly provided to us by Marcella
Carollo. The absolute cluster magnitudes range approximately from
$M_{\rm F606} \approx -7$ to $-16.5$ ($M_{\rm F606} \approx M_V +
0.2$, cf.~Holtzmann \etal 1995). Our best-fitting extinction-corrected
models for IC~342 indicate $M_{\rm F606} = -14.68$, under the
assumption that $D=1.8 \Mpc$. IC~342 has an enhanced star formation
rate (e.g., BFG97), consistent with Carollo's finding that this is
generally the case in galaxies with nuclear clusters for which $M_V
\lta -12$. Figure~\ref{f:carolsamp} shows that IC~342 clearly falls at
the high end of the luminosity distribution of the nuclear clusters
detected in other galaxies (and even more so if IC~342 is more distant
than assumed here). The results from our detailed study indicate that
the cluster in IC~342 is young, and stellar population models predict
that clusters fade in luminosity as they grow older.  Therefore, if
the nuclear clusters in this sample all have similar masses, then most
clusters would have to be older than the one in IC~342. To get a more
quantitative estimate of the age distribution we have used the
single-burst, Salpeter IMF stellar population models of Bruzual \&
Charlot (1993). Under the (very simplifying) assumption that all
observed nuclear clusters have the same mass as the one in IC~342, and
differ only in age, the age can be calculated directly from the
observed luminosity. The labels at the top of
Figure~\ref{f:carolsamp} indicate the ages thus calculated for each
luminosity bin. Under these simple assumptions most clusters are
inferred to be several Gyrs old, which seems entirely plausible.  This
is an interesting result, but is of course not a unique interpretation
of the observed cluster luminosity histogram. The faint clusters could
just be less massive than the cluster in IC 342, instead of being
older. This must in fact be the case for the very low-luminosity
clusters with $M_{\rm F606} > -9$, which are too faint to be faded
analogs of the IC~342 cluster for any assumed age.

The results from Figure~\ref{f:carolsamp} indicate that the nuclear
clusters in spiral galaxies need not all be as young as the one in IC
342. This is what one would expect if spiral galaxies do indeed have
only one episode of nuclear cluster formation, 
unless we are viewing all spiral galaxies at the same special
time in their evolution. Nonetheless, we may be seeing too many
young clusters for this scenario to be acceptable. If nuclear cluster
formation in spirals is a Poisson process with fixed probability per
unit time, we would expect to see clusters with ages of $10^7$, $10^8$
and $10^9$ years in the ratio $1:10:100$. The results in
Figure~\ref{f:carolsamp} suggest that this is not the case, and that
there is a higher fraction of young clusters than expected in this
scenario. This is consistent also with an inspection of just the four
closest spiral galaxies. Our Galaxy has a young central cluster that
is just $\sim 10^{6.5}$ years old (Krabbe
\etal 1995; Najarro \etal 1997), M31 and M33 have blue nuclei when
viewed at HST resolution that are quite possibly young star clusters
(Lauer \etal 1998), and IC~342 has a cluster that is $10^{6.8-7.8}$
years old.

To explain the high incidence of young star clusters we may have to
invoke scenarios in which nuclear cluster formation is a recurrent
process. Unfortunately, our observations do not provide any strong
arguments for or against this possibility. 
The equivalent width (EW) of the CO
bandhead at $2.29\mum$ provides some constraints on the presence of an
old underlying population (see also the discussion on continuous star
formation at the end of \S\ref{ss:popsynthesis}), but these are not
very strong. For example, if another instantaneous burst happened in
the nucleus of IC342 with the same stellar mass content, 
but $10^{9.5}$~yrs ago, 
its K-band light would be about 3 magnitudes fainter
than that of the currently observed young population (Bruzual \& Charlot 1993). 
Because of the small
contribution to the K-band continuum, the old cluster does not significantly
dilute the CO equivalent width: 
${\rm EW}_{\rm total} = {\rm EW}_{\rm young} - f({\rm EW}_{\rm young} -
{\rm EW}_{\rm old}]$, where $f$ is the fraction
of the total light contributed by the old cluster. For the
example above, this
yields only a 6\% decrease in the observed equivalent width, which
is not inconsistent
with the observations. Under these assumptions, the young cluster's 
mass and hence its mass-to-light ratio plotted in Figure~\ref{f:masstolight}
would be lower by a factor of two, still consistent with the models.

To make progress along these lines, it will be necessary to 
perform detailed studies of
the nuclear star clusters in many more spiral galaxies. A direct
determination of the age and mass distribution of the nuclear clusters
will be required to determine whether the formation of the nuclear
clusters in spiral galaxies is a recurring phenomenon or not, and how,
if at all, these clusters are related to the secular evolution of the
exponential bulges in which they reside.
    
 
\acknowledgments

We are grateful to David Zurek for assistance in the reduction and
analysis of the HST/WFPC2 images of IC 342, and to Marcella Carollo
for kindly making her data available to us in advance of publication.
We thank Bob Joseph for granting the observing time for this project
and the staff at the IRTF for their superb assistance at the
telescope. The stellar population models made available electronically
by Claus Leitherer and his collaborators and by Gustavo Bruzual and
Stephane Charlot were indispensable for the interpretation of our
data.



\clearpage

 
\ifsubmode\else
\baselineskip=10pt
\fi


\clearpage

 
\ifsubmode\else
\baselineskip=14pt
\fi

 
\newcommand{\figcapspectra}{{\bf Top:} Normalized CO-bandhead spectrum
of the nuclear cluster in IC~342, corrected to the heliocentric
system. {\bf Bottom:} Normalized spectra of four late-type template
stars, shifted horizontally to the redshift of IC 342. All spectra are
offset vertically for clarity. The heavy solid curve shows the best
Gaussian-convolved template fit to the galaxy spectrum, which has
$\sigma = 33 \kms$. By contrast, the heavy dashed curve shows the
(unacceptable) fit for a fixed dispersion of $\sigma = 62 \kms$. The
template spectra themselves correspond to $\sigma = 5.5 \kms$ (the
instrumental dispersion). A $\chi^2$ analysis shows that the nuclear
star cluster of IC 342 has $\sigma = (33 \pm 3)
\kms$.\label{f:spectra}}

\newcommand{\figcapimages}{{\bf (a; left)} IRTF/NSFCAM K-band image of 
the center of IC~342. The spatial resolution is 0.7\as\ (FWHM). A
logarithmic stretch was used to bring out the faint features. {\bf (b;
right)} HST/WFPC2 F606W (wide V-band) image of IC~342, on the same
scale as the K-band image, and with the same orientation. Again, a
logarithmic stretch was used to bring out the faint features. The
nuclear cluster stands out on a faint diffuse stellar background, and
is surrounded by a complicated pattern of dust absorption. There are
several fainter, off-nuclear star clusters. The clusters directly to
the North and South-South-West are fainter than the nuclear cluster by
$1.1$ and $0.7$ magnitudes, respectively. In the K-band image the
individual clusters are more difficult to discern due to the lower
spatial resolution. The morphology of the diffuse background light in
the K-band is more symmetric than in the V-band, due to the smaller
influence of dust absorption.\label{f:images}}

\newcommand{\figcapsurfbr}{Open dots show the V-band surface brightness 
profile (in mag/arcsec$^2$) of the nuclear star cluster of IC 342 as
inferred by ellipse fitting to the HST image shown in
Figure~\ref{f:images}b. The image is saturated in the central $r
\leq$ 0.08\as , which causes the flattening of the surface brightness profile 
at these radii. The two heavy curves are fits of a so-called nuker-law
parameterization. The lower curve is a fit to the data at $r \leq$
0.3\as , including the saturated pixels. The upper curve is a fit over
the same range, but excluding the saturated pixels at $r \leq$ 0.08\as .
In the absence of better data, both provide an acceptable model. Small
triangles indicate the azimuthally averaged profile of the PSF,
normalized so as to coincide with the galaxy profile at $r =$
0.08\as . The nuclear cluster is clearly more extended than the PSF,
and is therefore resolved.\label{f:surfbr}}

\newcommand{\figcapmasstolight}{Relation between $M/L_K$ and age 
for several model stellar clusters, as calculated with the
`Starburst99' software of Leitherer \etal (1999). The models have
solar metallicity ($Z = 0.02$). Different curves are models that
differ in the slope $\alpha$ and upper-mass cut-off $M_{\rm up}$ of
their IMF. Solid line: $\alpha = 2.35$ and $M_{\rm up} = 100 \Msun$;
long-dashed curve: $\alpha = 3.30$ and $M_{\rm up} = 100 \Msun$;
short-dashed curve: $\alpha = 2.35$ and $M_{\rm up} = 30 \Msun$.  All
models have a lower-mass IMF cut-off at $M_{\rm low} = 1 \Msun$, and
assume that the star formation occurs in a single burst. The feature
in the predicted $M/L_K$ at $\sim 10^{6.8}$ years is due to the
appearance of red supergiants, which is one of the more poorly
understood phases in the cluster evolution (cf.~Leitherer \etal 1999).
The heavy dashed horizontal line indicates the value $M/L_K = 0.047$
inferred in \S\ref{ss:dynmodel} for the nuclear star cluster of IC 342
under the assumption of an isotropic velocity distribution and $A_K =
0.45$. The models indicate that the cluster has $\log({\rm age}) \leq
7.7$.\label{f:masstolight}}

\newcommand{\figcapvminusk}{Relation between (unextinguished) 
$V-K$ and age for the same stellar population models as in
Figure~\ref{f:masstolight}. The observations for the nuclear cluster
of IC 342 give $(V-K)_{\rm obs} = 4.57$, indicating significant
extinction-induced reddening.\label{f:vminusk}}

\newcommand{\figcapAKage}{This figure illustrates how both 
cluster age and $A_K$ can be determined from the observed $(V-K)_{\rm
obs}=4.57$ and $M/L_K = 0.047 \times 10^{-0.4 (A_K - 0.45)}$. Each
panel corresponds to one of the three population models in
Figures~\ref{f:masstolight} and~\ref{f:vminusk}. The dotted curves
show the $A_K$ for which the predicted $V-K$ of the population model
matches the observations, and the solid curves show the $A_K$ for
which the predicted mass-to-light ratio matches the observations. The
points in the panels where the solid and dotted curves intersect
indicate combinations of age and $A_K$ that fit both observational
constraints. The allowed combinations have ages in the range
$10^{6.8-7.8}$ years, and $A_K$ in the range
$0.37$--$0.57$.\label{f:AKage}}

\newcommand{\figcapeqwidths}{Relation between the equivalent widths of 
three prominent near-IR absorption lines and age for the same stellar
population models as in Figure~\ref{f:masstolight}. The absorptions
lines are, from left to right, Si($1.59\mum$), CO($\nu =$6--3) at
$1.62\mum$ and CO($\nu =$2--0) at $2.29\mum$. The heavy dashed
horizontal lines indicate the values determined in BFG97. The errors
on these measurements are 5\%, 5\% and 10\%, respectively. These
measurements confirm the implication from the observed $M/L_K$ that
the nuclear star cluster has to be young.\label{f:eqwidths}}

\newcommand{\figcapcontburst}{{\bf (a; left)} Relation between $M/L_K$ 
and age for several model stellar clusters. Models and curves are as
in Figure~\ref{f:masstolight}, but now assume a constant star
formation rate since $t=0$, instead of an instantaneous burst. The
heavy dashed horizontal line indicates the value $M/L_K = 0.047$
inferred in \S\ref{ss:dynmodel} for the nuclear star cluster of IC 342
under the assumption of an isotropic velocity distribution and $A_K =
0.45$. {\bf (b; right)} Relation between the equivalent width of the
CO($\nu =$2--0) bandhead at $2.29\mum$ and age for the same models as
in the left panel. The heavy dashed horizontal lines indicates the
value determined in BFG97, which has a measurement error of $10$\%.
The large observed equivalent width rules out models with a constant
star formation rate.\label{f:contburst}}

\newcommand{\figcapcarolsamp}{Histogram of nuclear stellar cluster 
luminosities for the combined galaxy sample of Carollo \etal (1997)
and Carollo, Stiavelli \& Mack (1998). Only galaxies with detected
nuclear clusters are included. The absolute magnitude along the
abscissa, $M_{\rm F606}$, is in the synthetic instrumental magnitude
system of the WFPC2 (Holtzmann \etal 1995) for which typically,
$M_{\rm F606} \approx M_V + 0.2$. Our results for IC 342 indicate an
(extinction corrected) $M_{\rm F606} = -14.68$, as indicated by the
arrow. IC 342 is itself not included in the sample used to construct
the histogram. The labels at the top indicate $\log({\rm age})$ for
each bin, using a simple model that assumes that all clusters are
identical except for their age. The luminosities were calculated from
the stellar population models of Bruzual \& Charlot (1993), and are
normalized to $\log({\rm age}) = 7.4$ for IC 342.\label{f:carolsamp}}

 
\ifsubmode
\figcaption{\figcapspectra}
\figcaption{\figcapimages}
\figcaption{\figcapsurfbr}
\figcaption{\figcapmasstolight}
\figcaption{\figcapvminusk}
\figcaption{\figcapAKage}
\figcaption{\figcapeqwidths}
\figcaption{\figcapcontburst}
\figcaption{\figcapcarolsamp}
\clearpage
\else\printfigtrue\fi

\ifprintfig
 
 
\clearpage
\begin{figure}
\epsfxsize=12.0truecm
\centerline{\epsfbox{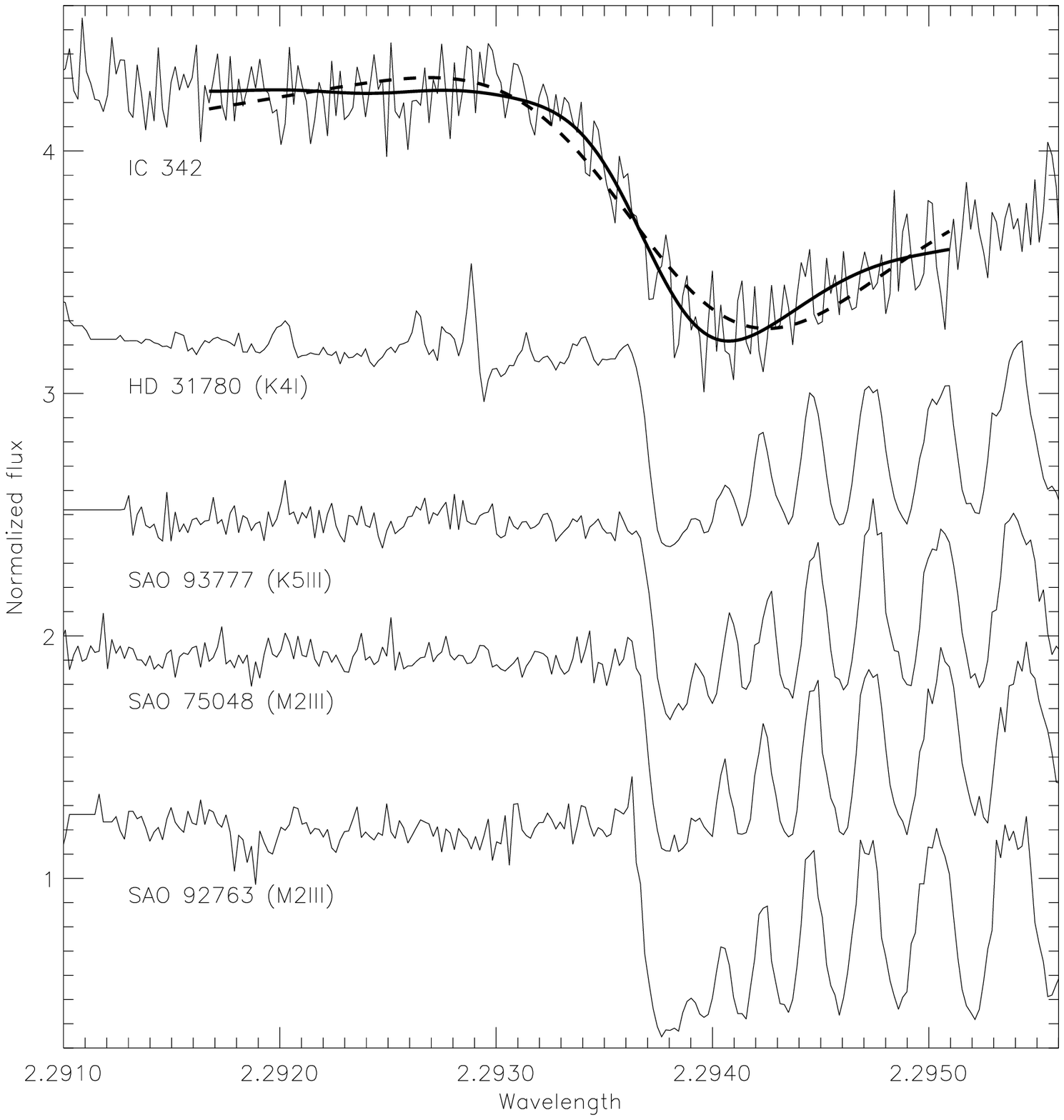}}
\ifsubmode
\vskip3.0truecm
\setcounter{figure}{0}
\addtocounter{figure}{1}
\centerline{Figure~\thefigure}
\else\vskip-0.3truecm\figcaption{\figcapspectra}\fi
\end{figure}
 

\clearpage
\begin{figure}
\noindent
\begin{minipage}[b]{0.475\hsize}
\epsfxsize=0.98\hsize
\epsfbox{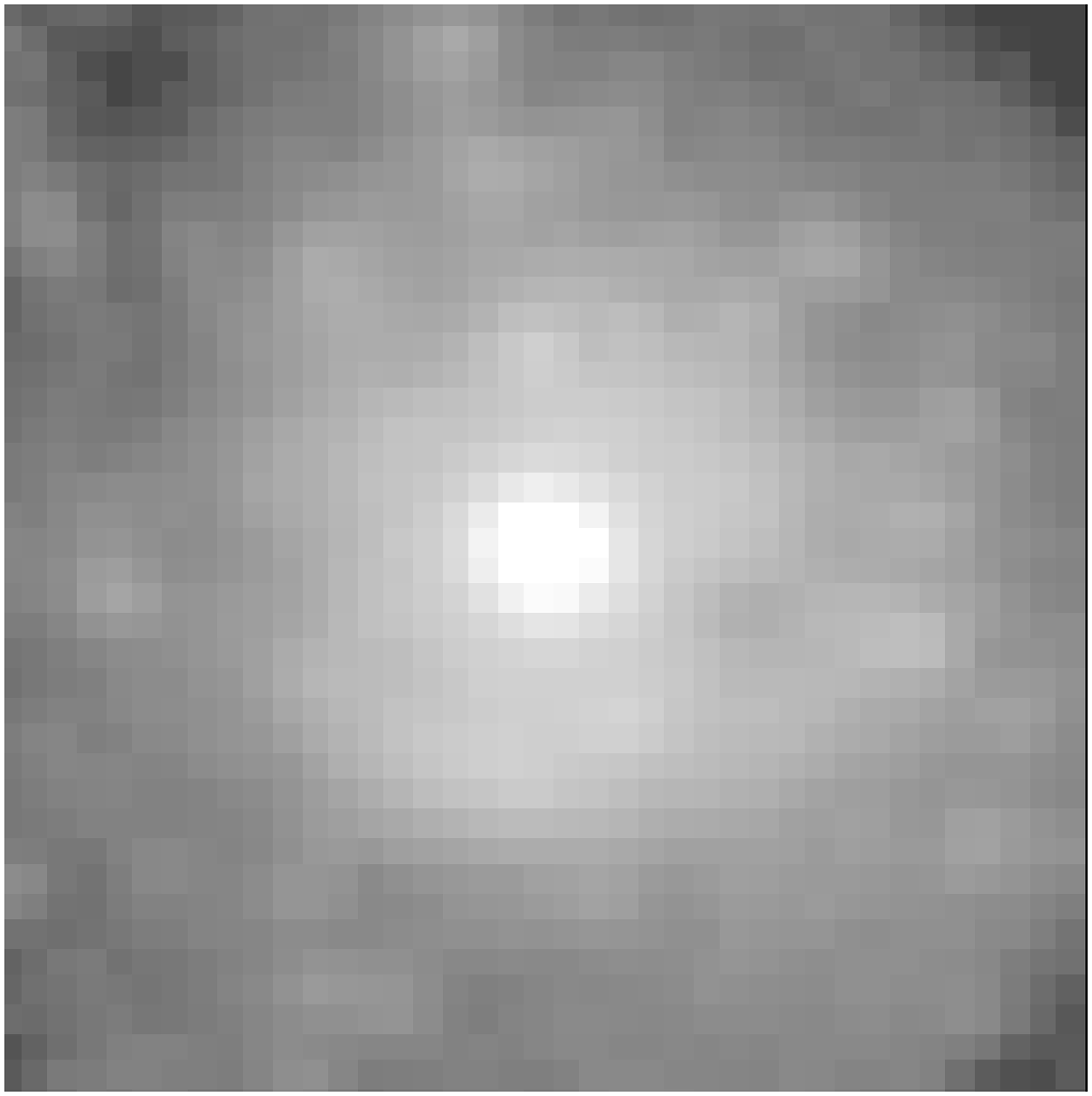}
\end{minipage}
\hfill
\begin{minipage}[b]{0.475\hsize}
\epsfxsize=0.98\hsize
\epsfbox{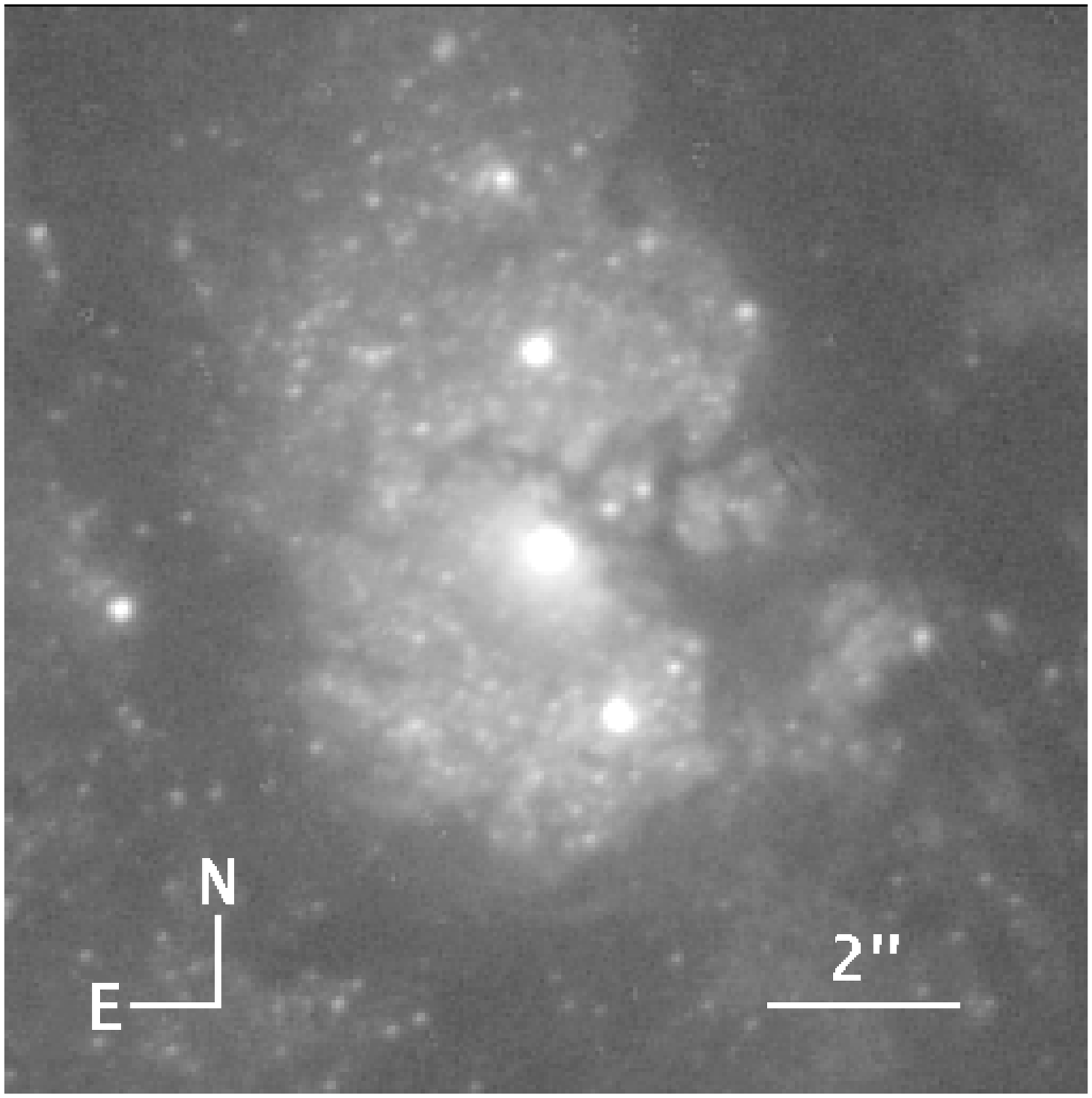}
\end{minipage}
\ifsubmode
\vskip3.0truecm
\addtocounter{figure}{1}
\centerline{Figure~\thefigure}
\else\figcaption{\figcapimages}\fi
\end{figure}


\clearpage
\begin{figure}
\centerline{\epsfbox{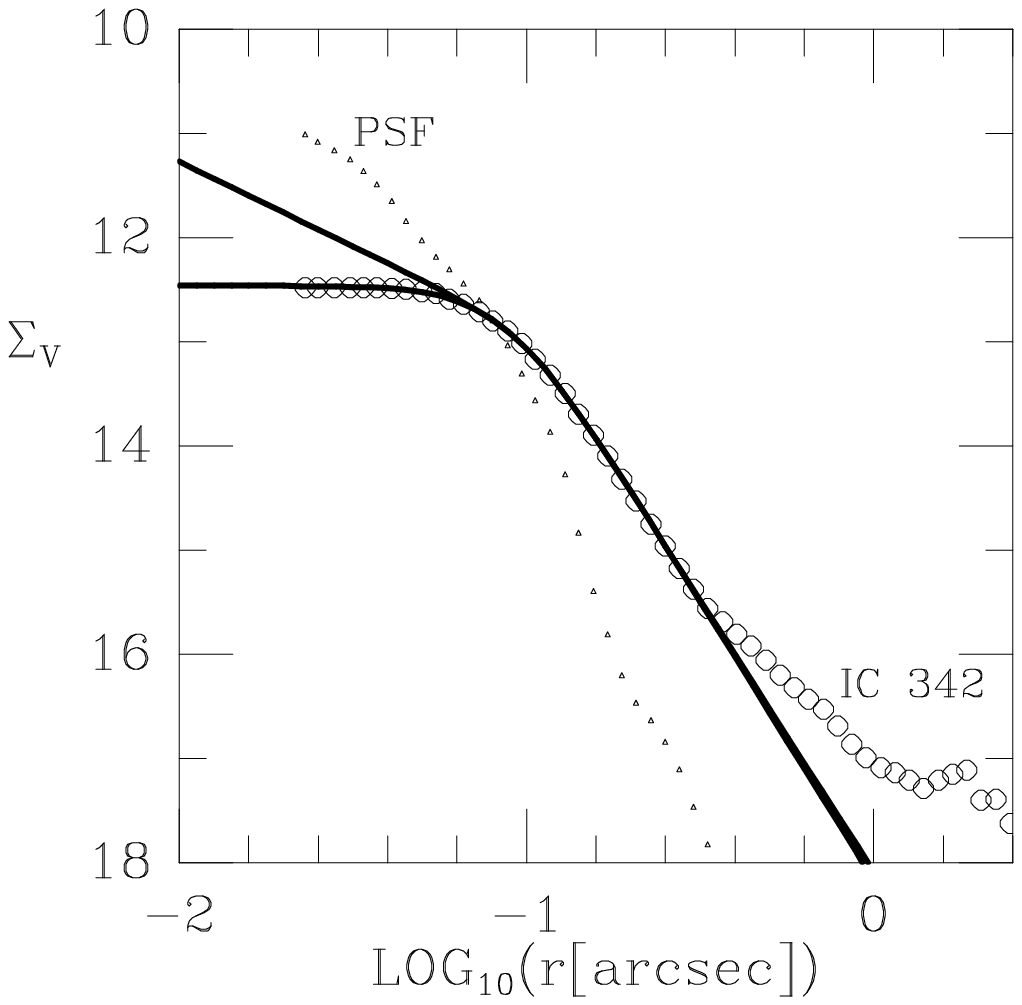}}
\ifsubmode
\vskip3.0truecm
\addtocounter{figure}{1}
\centerline{Figure~\thefigure}
\else\figcaption{\figcapsurfbr}\fi
\end{figure}


\clearpage
\begin{figure}
\centerline{\epsfbox{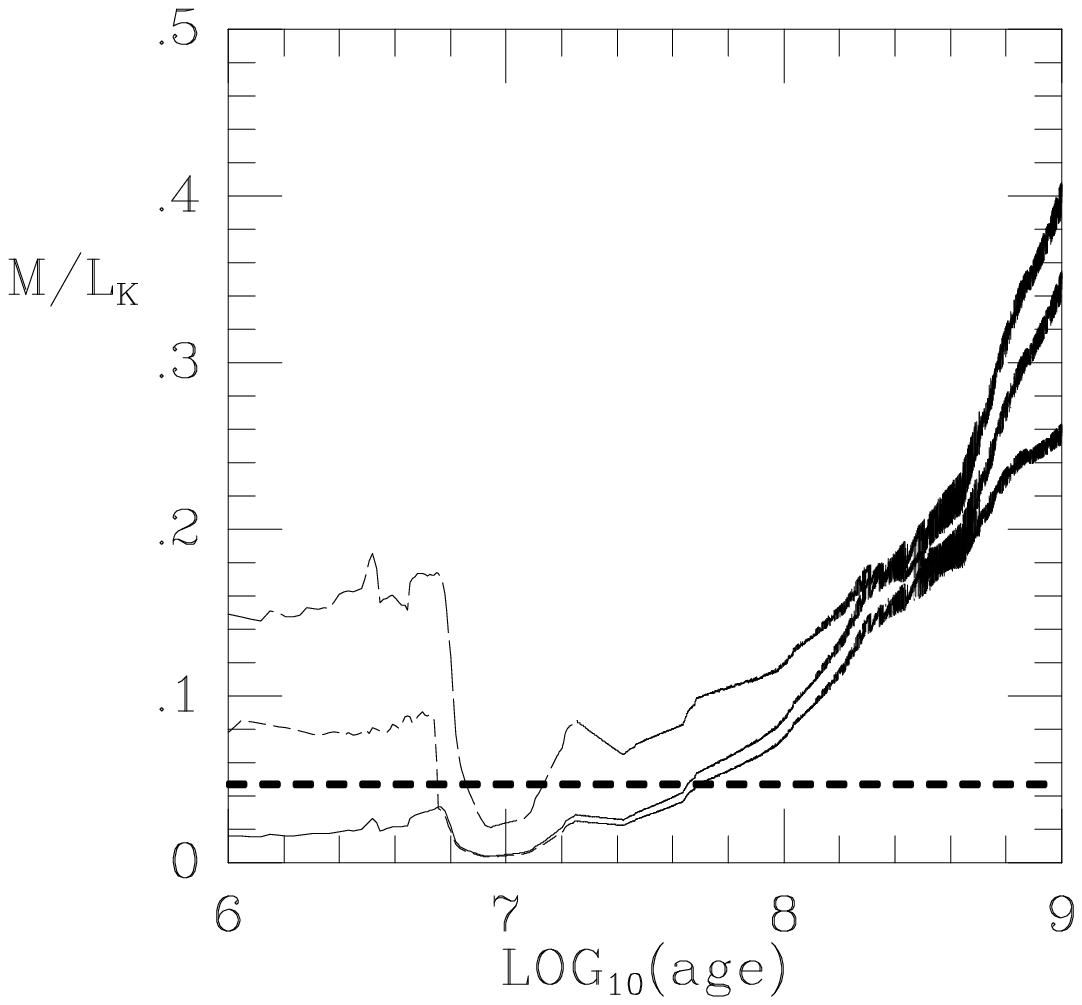}}
\ifsubmode
\vskip3.0truecm
\addtocounter{figure}{1}
\centerline{Figure~\thefigure}
\else\figcaption{\figcapmasstolight}\fi
\end{figure}


\clearpage
\begin{figure}
\centerline{\epsfbox{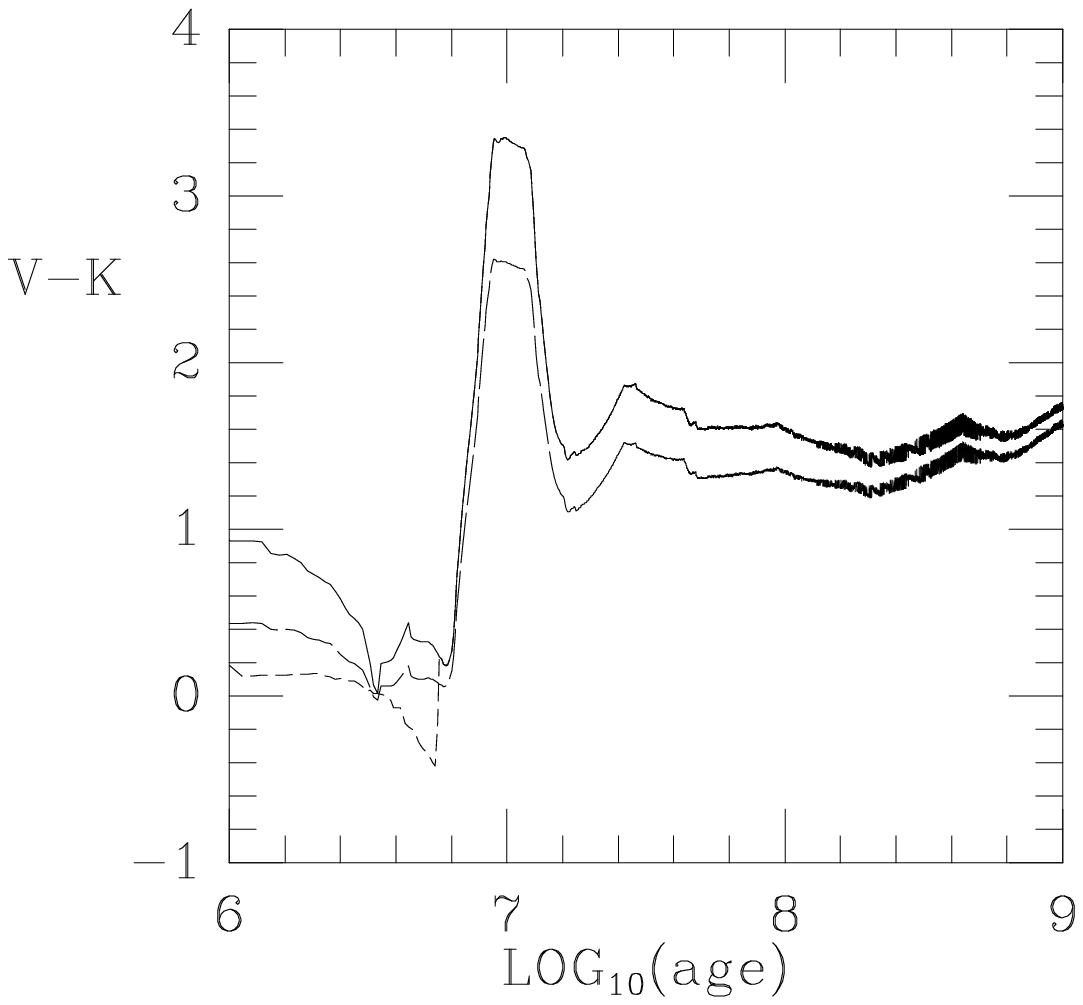}}
\ifsubmode
\vskip3.0truecm
\addtocounter{figure}{1}
\centerline{Figure~\thefigure}
\else\figcaption{\figcapvminusk}\fi
\end{figure}


\clearpage
\begin{figure}
\centerline{\epsfbox{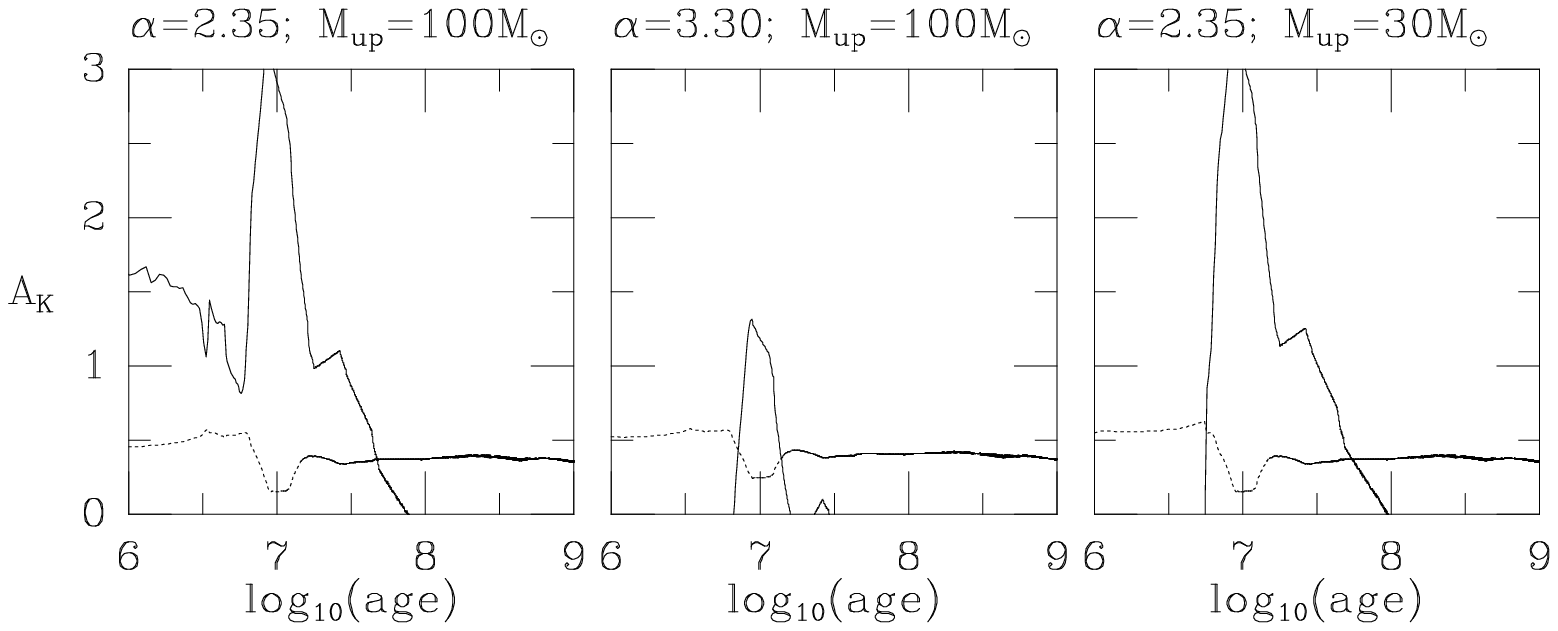}}
\ifsubmode
\vskip3.0truecm
\addtocounter{figure}{1}
\centerline{Figure~\thefigure}
\else\figcaption{\figcapAKage}\fi
\end{figure}


\clearpage
\begin{figure}
\centerline{\epsfbox{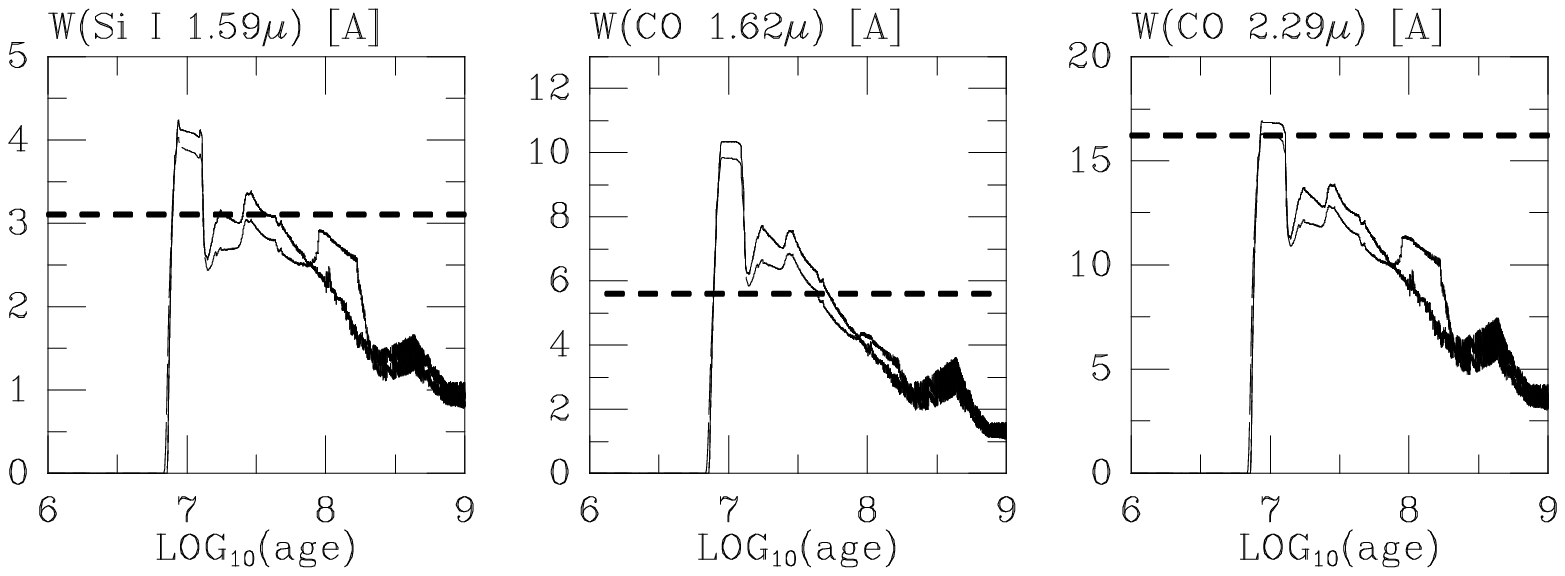}}
\ifsubmode
\vskip3.0truecm
\addtocounter{figure}{1}
\centerline{Figure~\thefigure}
\else\figcaption{\figcapeqwidths}\fi
\end{figure}


\clearpage
\begin{figure}
\centerline{\epsfbox{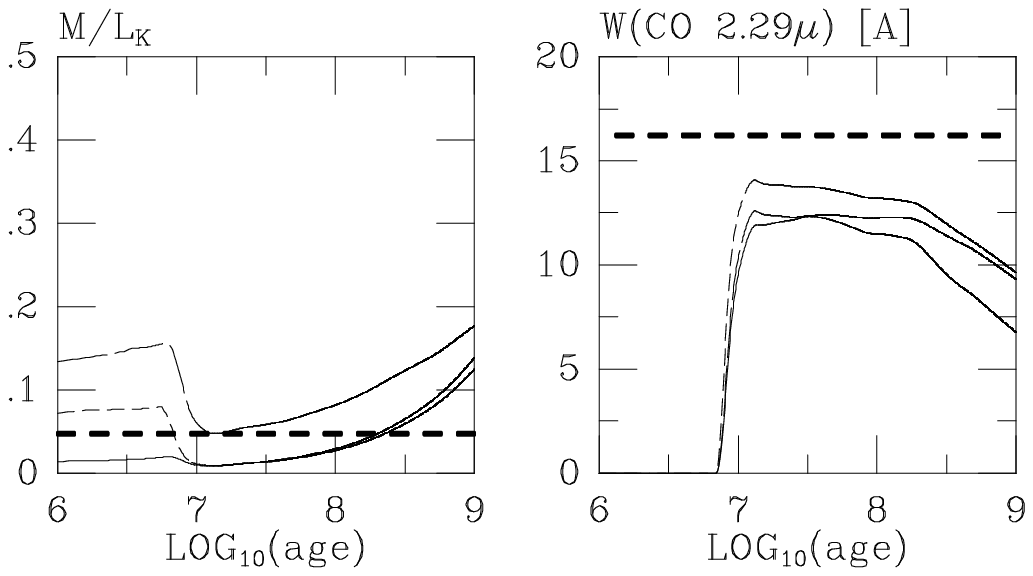}}
\ifsubmode
\vskip3.0truecm
\addtocounter{figure}{1}
\centerline{Figure~\thefigure}
\else\figcaption{\figcapcontburst}\fi
\end{figure}


\clearpage
\begin{figure}
\centerline{\epsfbox{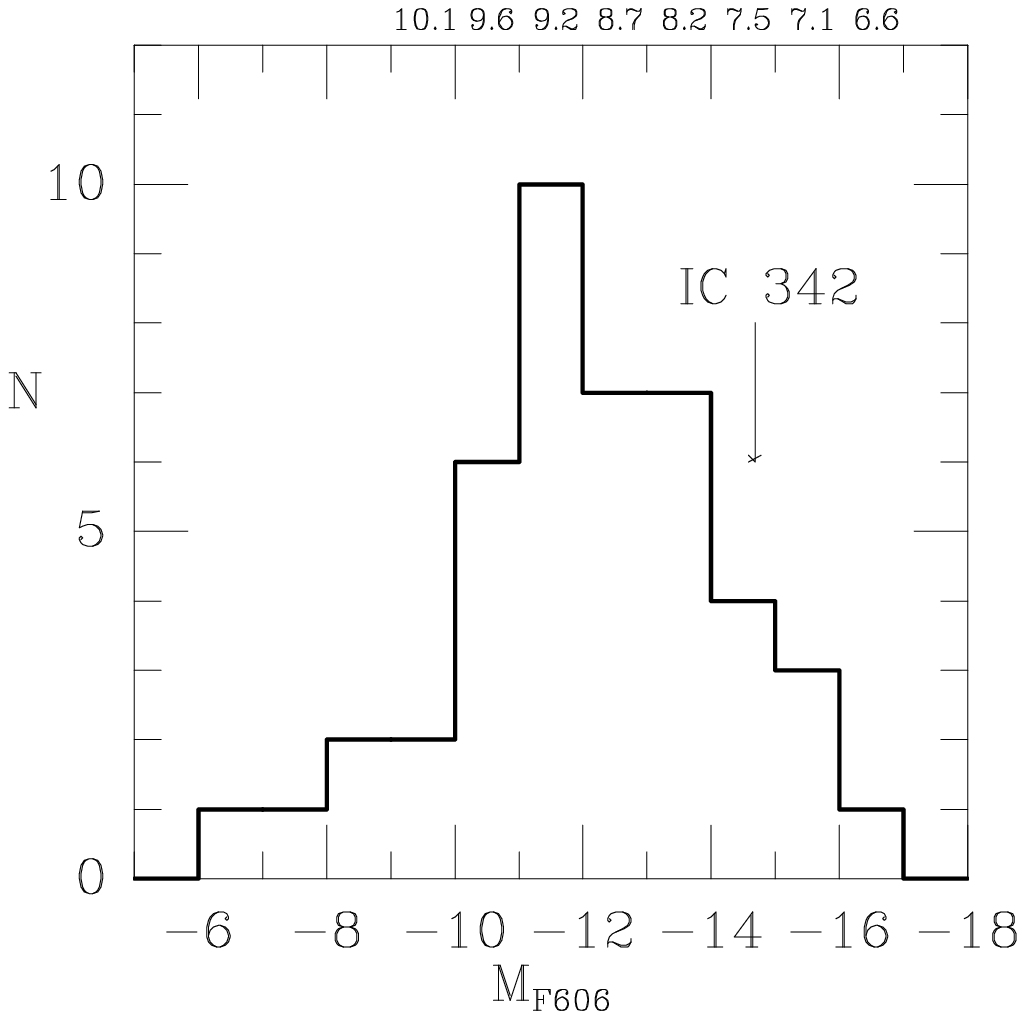}}
\ifsubmode
\vskip3.0truecm
\addtocounter{figure}{1}
\centerline{Figure~\thefigure}
\else\figcaption{\figcapcarolsamp}\fi
\end{figure}

 
\fi

 
 


 
\end{document}